# Fitting the structural relaxation time of glass-forming liquids: single- or multi-branch approach?

Lianwen Wang

Institute of Materials Science and Engineering and MOE Key Laboratory for Magnetism and Magnetic Materials, Lanzhou University, Lanzhou 730000, P. R. China.

Email: lwwang@lzu.edu.cn

Glass transition and the dynamics of glass-forming liquids[1-3], from organic, oxide, metallic glasses to polymers, are of the central interests of researchers in materials science[4-6], cryobiology[7], geology[8,9], so on and so forth. A main challenge lies in the understanding of the super-Arrhenius temperature dependence of the structural relaxation time $\tau$ (or viscosity $\eta=\tau G_\infty$ where $G_\infty$ is the instantaneous shear modulus[3]) near glass transition temperature, $T_g$.

It is known that at high temperatures, e.g. above the melting point $T_m$, the temperature dependence of the structural relaxation time of a liquid is Arrhenius[10,11]:

$$\tau = \tau_0 \exp\left(\frac{E}{kT}\right), \qquad (1)$$

where $\tau_0$ is a material dependent pre-exponential factor and $E$ the activation energy. However, with temperature decreasing, the relaxation time of glass-forming liquids will increase dramatically by order of magnitudes within several tens of degrees above $T_g$, departing significantly from the Arrhenius law[1-3]. A major confusion in understanding glass transition has been i) how to describe the temperature dependence of the relaxation time of glass-forming liquids and ii) what is the origin of this super-Arrhenius behavior?

As early as in 1920s H. Vogel, G. Tammann and W. Hess, and G. S. Fulcher[12,13] proposed, independently, a three-parameter empirical equation for oxide glass melts:

$$\tau = A\exp\left(\frac{B}{T-T_0}\right), \qquad (2)$$

where $A$, $B$ and $T_0$ are material dependent constant. Because of its simplicity Eq. 2, known as the VTF equation, has ever since been widely applied[1-3] even though it fails for some materials, with the origin of the super-Arrhenius behavior remains unresolved.

Nonetheless, the failure of the VTF equation is unneglectable and is obvious: analysis of measured viscosity data for oxide[12], organic[11,14-17] and metallic[18] glass melts *all* showed that the VTF equation failed for a *full* temperature range, i.e. from above $T_m$ down to $T_g$.

Rather, Macedo and Litovitz[19], Battezzati[18], and Richert *et. al.*[11,20] found that, in a full temperature range, the relaxation time of glass-forming liquids should be fitted by a three-branch method, namely a high temperature branch, a low temperature branch, and an intermediate branch connecting the high and the low temperature branches. The above authors agreed that the high temperature branch was Arrhenius and the intermediate branch VTF. As to the low temperature branch, Macedo and Litovitz[19] and Battezzati[18] found that it should be Arrhenius. Although Richert *et. al.*[11,20] fitted the low temperature branch with a VTF equation, the Arrhenius nature of their measured data[11], and in other measurements e.g. Ref. 16, at temperatures near $T_g$ was indeed obvious. As such, the discrepancies in fitting the low temperature branch should come from the critical issue of how to fit, meaningfully, the intermediate region between the high and the low temperature Arrhenius branches.

Recently, this author worked out two Arrhenius equations for the high and the low temperature branches[21]. Relaxations in the low temperature branch were cooperative and showed different slopes from the high temperature non-cooperative branch. Here it is shown that the gradual change between the high and the low temperature Arrhenius branches, represented by Eq. 7 and Eq. 8 in Ref. 21, was caused by the gradual increase of atomic cooperativity in structural relaxations:

$$\tau = \frac{\tau_0}{C_m}\left\{1+(1-C_m)\left[\frac{1/[(1-C_m)N]}{C_m}\right]^{(1-C_m)N}\right\} \quad (C_m \leq 1), \quad (3)$$

where $C_m$ is the the possibility that an atom could migrate (or the concentration of migration atoms) and $N$ the number of atoms involved in atomic cooperativity[21]. With temperature descending, cooperative relaxation, hence departure from the Arrhenius law, occurred when $C_m$ became less than unity. Further decreasing the temperature, $C_m$ decreased exponentially to nearly zero and the degree of cooperativity in relaxation approached its upper limit represented by the low temperature Arrhenius branch. Detailed explanations of Eq. 3 will be given in Ref. 22.

In Fig. 1 reported structural relaxation data of Glycerol[23-25] were plotted in a $\log\tau$-($T_g/T$) scale and were fitted by using Eq. 1 and Eq. 3. Measurement inaccuracies should be taken into account when judging the quality of present fittings to measured relaxation data. In comparison with the three-branch method by Macedo and Litovitz[19], Battezzati[18], and Richert *et. al.*[11,20], a two-branch method was used here, i.e.

an Arrhenius equation for $C_m>1$ and a gradual-changing branch for $C_m<1$. A significant merit of this method is that the gradual changes in the low temperature branch and the turning point between the two branches were meaningfully and quantitatively given.

Still there were other attempts to fit the viscosity data of glass-forming liquids with a single formula in a full temperature range, e.g. the model of Avramov and Milchev[26-28] and that of Mauro *et. al.*[29]. However when tested with measured relaxation data, the model of Mauro *et. al.* had not showed significant superiority over the VTF equation[30] and the model of Avramov and Milchev was criticized[31].

To sum up, in fitting the viscosity of glass-forming liquids, the single-branch approach has a tradition traced back to 1920s, but does not produce convincing results with clear physics. This note is trying to recall the attention of researchers in this field to the possibilities of the multi-branch approach.

**Figure Captions**

**Figure 1** A two-branch fitting to measured structural relaxation data[23-25] of Glycerol. Dashed lines are two Arrhenius fittings for the high and the low temperature branches, respectively. With temperature decreasing, departure from the high temperature Arrhenius branch occurred when the possibility that an atom could migrate, $C_{\mathrm{m}}$, became less than unity so that cooperativity was needed in relaxation (solid line). $C_{\mathrm{m}}$ decreased exponentially to zero with temperature descending hence the degree of atomic cooperativity approached its upper limit, as was represented by the low temperature Arrhenius branch.

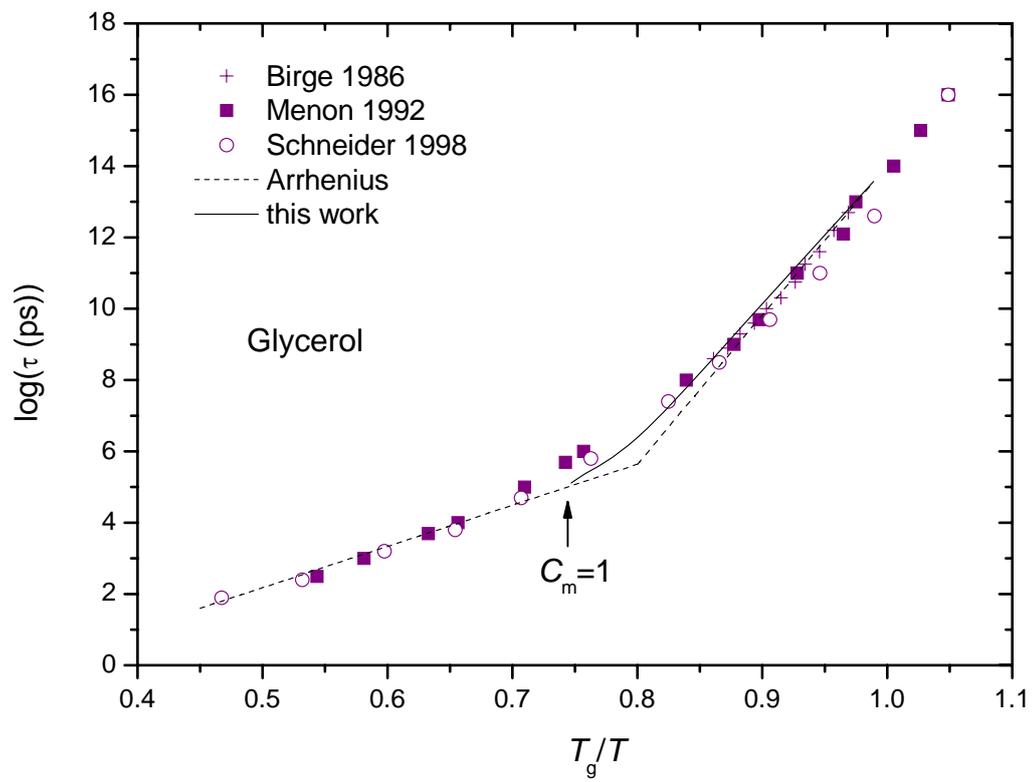

**Figure 1/Wang**